\documentstyle{article}  
\newtheorem{th}{Theorem}  
\newtheorem{lm}{Lemma}
\newtheorem{df}{Definition}   
\newtheorem{pr}{Proposition}   
\newtheorem{as}{Assumption}  
\newcommand{\bth}{\begin{th}}
\newcommand{\eth}{\end{th}}
\newcommand{\blm}{\begin{lm}}  
\newcommand{\elm}{\end{lm}}
\newcommand{\bdf}{\begin{df}}  
\newcommand{\edf}{\end{df}} 
\newcommand{\bpr}{\begin{pr}}
\newcommand{\epr}{\end{pr}}  
\newcommand{\bas}{\begin{as}}
\newcommand{\eas}{\end{as}}

\newcommand{\bit}{\begin{itemize}}
\newcommand{\eit}{\end{itemize}\par\noindent}
\newcommand{\beq}{\begin{equation}} 
\newcommand{\eeq}{\end{equation}\par\noindent}
\newcommand{\beqa}{\begin{eqnarray*}}
\newcommand{\eeqa}{\end{eqnarray*}\par\noindent}
\newcommand{\beqn}{\begin{eqnarray}}
\newcommand{\eeqn}{\end{eqnarray}\par\noindent}  
 
\newfam\msbfam
\font\tenmsb=msbm10                     \textfont\msbfam=\tenmsb
\font\sevenmsb=msbm7            \scriptfont\msbfam=\sevenmsb
\font\fivemsb=msbm5                     \scriptscriptfont\msbfam=\fivemsb
\def\Bbb{\fam\msbfam \tenmsb}

\title{\bf A (classical) Representation for a Spin-S Entity as a\\ compound system in {\sf
IR}$^3$   consisting of\\ 2S Individual Spin-1/2 Entities.}
\date{}
\author{Bob Coecke\thanks{Free University of Brussels (VUB), Department of Mathematics, Pleinlaan
2, B-1050 Brussels; bocoecke@vub.ac.be\,.}} 

\begin{document}  

\maketitle
\vspace{-6cm} 
\noindent
\centerline{Appeared in Foundations of Physics {\bf 28}, 1347--1365 (1998) --- submitted in
1996;}

\smallskip\noindent\centerline{(!) note that the published title doesn't have ``(classical)'' in
it.}   
\vspace{5.5cm}
\noindent  
\begin{abstract}
\noindent  
We generalize the results of \cite{coe95a} for the coherent states of
a spin-1 entity to spin-S entities with $S>1$ and to non-coherent spin states: through the 
introduction of 'hidden correlations' (see \cite{HC:FPL}) we introduce a representation for a spin-S
entity as a compound system consisting of $2S$ 'individual' spin-1/2 entities, each of them
represented by a 'proper state', and such that we are able to consider a measurement on the spin-S
entity as a measurement on each of the individual spin-1/2 entities.  If the spin-S entity is in a
maximal spin state, the 
$2S$ individual spin-1/2 entities behave as a collection of indistinguishable but separated
entities.  If not so, we have to introduce the same kind of hidden correlations as
required for a hidden correlation representation of a compound quantum systems described by a
symmetrical  superposition.  Moreover, by applying the Majorana representation of \cite{maj32} and
Aerts' representation for a spin-1/2 entity of \cite{aer86}, this hidden correlation representation
yields a classical mechanistic representation of a spin-S entity in 
${\Bbb R}^3$.    
   
\end{abstract}  


\section{Introduction.}

In this paper, we represent a spin-S entity
as a compound system consisting of $2S$ {\it individual} spin-1/2 entities between which there exist
the kind of correlations that we have introduced in \cite{HC:FPL}, called {\it hidden correlations}. 
Such a hidden correlation representation implies that a measurement on the spin-S entity can be
represented as 2S measurements on the individual spin-1/2 entities, in the sense that a
measurement on one of the individual spin-1/2 entities induces a change of the {\it proper state} of
the remaining not yet measured individual spin-1/2 entities.  
A more general conceptual framework in which we can consider these hidden 
correlation representations has already been presented in \cite{HC:FPL}.  However, without
having to repeat to much of this framework we have made this paper formally
self-consistent\footnote{Since in this paper the focus is on purely
formal results of which some do also stand without a preferred approach to quantum measurements, we
will avoid as much as possible to refer to any interpretative (and thus speculative) framework, in
contrary to
\cite{HC:FPL}.  However, it is indeed possible to embed the results of this paper within the
conceptual framework of
\cite{HC:FPL} in a straightforward way.  Necessary remarks on this embedment will be included in this
paper as footnotes.}. The concepts of {\it
individual entities} and {\it proper states} introduced in
\cite{HC:FPL} will be used in this paper (without any further explanation) in order to
distinguish between the {\it proper states} of the {\it  individual entities} in the compound system,
i.e., the individual spin-1/2 entities, and the {\it state} of the {\it entity} that corresponds with
the compound system itself, i.e., the spin-S entity\footnote{However, there are more profound reasons
besides clarity to distinguish between states and proper states, and entities and individual
entities.  For a detailed discussion on this matter we refer to
\cite{HC:FPL}.  If we consider the results that we will obtain in this paper relative to the more
general framework introduced in \cite{HC:FPL}, it follows that from a purely formal point of view the
introduction of the notion of proper state is not really required
since for the representation introduced in this paper it is possible to represent the proper states of
the individual spin-1/2 entities as the states of a spin-1/2 entity.  In the language
of \cite{HC:FPL} this means that the measurements on spin-S entities do not 
include a hard act of creation.}.   

As mentioned in the abstract, we generalize the results of
\cite{coe95a} for the {\it coherent states} of a spin-1 entity to spin-S entities with $S>1$ and to
{\it non-coherent spin states}.  In fact, because of its simple geometry, the spin-1 representation in
\cite{coe95a} illustrates in a very simple way the true nature of the representation introduced in
this paper.   Although the
representation is also valid for non-coherent spin states, we first deduce all results for the
specific case of coherent spin states.  There are two main reasons for this.  At first, it makes all
proofs and derivations extremely transparent and natural, with the state space
as the primal mathematical object. Secondly, there are
serious reasons to think that the only real physical spin states are the coherent ones, which makes
this {\it partial representation} for the coherent spin states the essential one (see for example
\cite{aab94}). 

The existence of the representation introduced in this paper can be brought back to the
existence of the following three {\it sub-representations}:
\par
\smallskip 
\noindent
{\it 
{\bf 1.} It is possible to represent the states of the spin-S entity as $2S$ spin-1/2 states in a one
to one way by applying Majorana's representation of
\cite{maj32}.   {\bf 2.} It is possible to represent the {\it transition probabilities} of spin-S
entities in
$\,\bigl(\otimes{\Bbb C}^2\bigr)^{2S}\cong{\Bbb C}^{4S}$, i.e., in the
tensor product in which compound systems of $2S$ individual spin-1/2 entities are described.  This is
proved in section
\ref{sec:sym} of this paper.   {\bf 3.} Every measurement on a compound system described
in the tensor product of finite number of Hilbert spaces can be represented as a collection
of measurements on the individual entities in this compound system, each of them represented by a
proper state, and on which we introduce hidden correlations (this has been proved in
\cite{HC:FPL}, where we have made an explicit construction of a hidden correlation representation
for compound quantum systems described in a tensor product of Hilbert spaces).  We perform this last
step of the representation in section \ref{rep.HC}.}
\par 
\smallskip  
\noindent
These three sub-representations together yield the representation of a spin-S entity as 2S 
individual spin-1/2 entities.  Moreover, we can go one step further. 
If we combine our representation with 
Aerts' classical mechanistic model system
of \cite{aer86} for a spin-1/2 entity in the three dimensional space of reals we obtain a classical
mechanistic model system in the three dimensional space of reals for all spin-S entities (this is
discussed in section \ref{sec:HC-HF} of this paper).  
 
As already mentioned above, one of the main ingredients in this paper is the Majorana representation. 
In his famous paper of 1932, Majorana showed that the study of the angular momentum
$J$ of a quantum entity in a varying magnetic field can be reduced to the study of $2J$ angular
momenta with value 1/2, i.e., we only have to consider $2J$ representative points on a 
sphere.  Although this particular representation proved its advantage in experimental
applications\footnote{It enables the experimentalist to 'visualize' and 'manipulate' the angular
momenta of quantum entities within our three dimensional macroscopical experimental context (see
for example \cite{blo45}).}, to our knowledge, it has never been used in issues that deal with the
fundamental interpretation of the mathematical structures encountered in the quantum framework, and in
particular in quantum measurement theory.   We also remark that all the to the present author's
knowledge known proofs of the Majorana theorem deduce the existence of
a representation as spin-1/2 momentum operators from the algebraic properties of the momentum 
operators, 
and thus, one never uses a representation of the spin-S states as spin-1/2 states through an explicit
canonical embedment.  A reason for this is probably that most authors even seem to refuse to attach a
definite interpretative meaning to the representation\footnote{Except for Schwinger in \cite{sch77},
but his interpretation is purely deduced from group theoretic concepts, and thus heading in a
completely different direction than the approach elaborated in this paper.} (see for example
\cite{blo45}, 7th line of the third paragraph).  Therefore we think that the specific proof of the Majorana
theorem in this paper contributes to the understanding and transparency of the Majorana theorem
itself.

\section{A representation for coherent spin states.}

As we mentioned in the introduction, we proceed in three steps: in the first subsection we present 
the Majorana representation; in the next one we construct an explicit procedure to relate the
transition probabilities of the spin-S states to the proper states of $2S$ individual spin-1/2
entities in a compound system; finally, due to the specific nature of this procedure, we are
able to introduce deterministic correlations between the individual spin-1/2 entities in order to
consider a measurement on a spin-S entity as 2S measurements on the individual spin-1/2 entities in
the compound system, where the initial proper states of the individual entities is given by the
Majorana representation.

\subsection{Decomposition of the spin-S state space into spin-1/2
states.}\label{coh.Maj.}
 
First we study the transition probabilities of
measurements on spin-1/2 entities, and then we do the same for spin-S entities. This will enable us to
introduce the Majorana representation in an explicit way for coherent spin-S states.  In fact, we
consider a transition probability  as it is defined in \cite{acc82}
or \cite{mie68}, i.e., if $\Sigma$ is the state space of a physical entity, 
then the transition probability is a map
$P:\Sigma\times\Sigma\rightarrow [0,1]$ which is such that for a fixed state $p$, $P(p,q)$ is the
probability to obtain as the outcome of a
measurement the one that corresponds\footnote{Of course, this only makes sense if
we are able to identify the outcomes of the measurements with certain states of the entity.  This is
definitely the case if we follow  Pauli's definition of a 
measurement of the first kind (see
\cite{pau} and
\cite{pir}) where we suppose that after a measurement on the entity, its state has changed into a state
that corresponds with the obtained outcome.} with the state $q$.  In the case of quantum entities the transition
probability is for every measurement given by the square of the Hilbert space in-product of the unit
vectors which represent the states.  The states which represent the possible
outcomes of a measurement will be called {\it outcome states}.

\subsubsection{The transition probability for spin-1/2 states.}

In this section, we represent the states of 
a spin-1/2 entity on the Poincar\'e sphere ${\cal S}$.  First we need to calculate the transition
probability for spin-1/2 entity.    The transition probability depends only on the relative position of
the Stern-Gerlach apparatus\footnote{A Stern-Gerlach apparatus might be called the standard
measurement tool to measure the spin of an entity.  Its relevant parameters are a direction and a
sense in space.} in which we prepare the entity and the second Stern-Gerlach with which we measure the
spin. We represent such a measurement by the Euler angles 
$\alpha,\beta,\gamma$, and denote it as
$e_{\alpha,\beta,\gamma}$. If the initial state corresponds with a spin quantum number $m=+{1\over 2}$
we denote it as $p_{+}^0$, and if it corresponds with a spin quantum number
$m=-{1\over 2}$ we denote it as $p_{-}^0$. We represent $p_{+}^0$ by the
vector $\psi_{+}^0=(1,0)\in{\Bbb C}^2$ and $p_{-}^0$ by
$\psi_{-}^0=(0,1)\in{\Bbb C}^2$.  The eigenvectors corresponding to a
measurement $e_{\alpha,\beta,\gamma}$ are the same as the ones we obtain when
we rotate the initial states by an active rotation characterized by the Euler
angles $\alpha,\beta,\gamma$. This active rotation is represented by the
unitary operator acting on ${\Bbb C}^2$ that corresponds with the
following matrix (the derivation of this matrix can be found in \cite{wig59}): 
\beq\label{eq:1}
M_{\alpha,\beta,\gamma}=
\pmatrix{
e^{-I{\alpha +\gamma\over 2}} cos{\beta\over 2}  & -e^{-I{\alpha -\gamma\over 2}}sin{\beta\over2}\cr 
e^{I{\alpha -\gamma\over 2}}sin{\beta\over 2}  & e^{I{\alpha +\gamma\over 2}}cos{\beta\over 2}\cr 
}
\eeq   
Thus, for the measurement
$e_{\alpha,\beta,\gamma}$ we have a set of outcome states represented by the
following eigenvectors:
\beqn
\label{eq:2a}
\psi_+^{\alpha,\beta,\gamma}
&=&
M_{\alpha,\beta,\gamma}\psi_{+}^0=
\bigl(e^{-I{\alpha\over 2}}cos{\beta\over 2} , e^{I{\alpha\over 2}}sin{\beta\over 2}
\bigr)e^{-I{\gamma\over 2}}
\\ \label{eq:2b}
\psi_-^{\alpha,\beta,\gamma}
&=&
M_{\alpha,\beta,\gamma}\psi_{-}^0=
\bigl(-e^{-I{\alpha\over 2}}sin{\beta\over 2} , e^{I{\alpha\over 2}}cos{\beta\over 2}
\bigr)e^{I{\gamma\over 2}}
\eeqn
The vectors in
eq.(\ref{eq:2a}) and eq.(\ref{eq:2b}) that correspond to different values of $\gamma$ (for fixed
$\alpha$ and $\beta$) represent the same states. As a consequence, we omit
the superscript $\gamma$ in the notations for the vectors and the
measurements. We represent the states corresponding to the vectors in eq.(\ref{eq:2a})
and eq.(\ref{eq:2b}) respectively by $p_+^{\alpha,\beta}$ and $p_-^{\alpha,\beta}$. 
Thus we have  $p_+^{0,0}=p_+^0$ and $p_-^{0,0}=p_-^0$. We also have:
\beq\label{eq:3}
p_+^{\alpha,\beta}=p_-^{\alpha+\pi,\pi-\beta} 
\eeq 
We denote the
transition probability to obtain a state
$p_+^{\alpha,\beta}$ in a measurement
$e_{\alpha,\beta}$ on an entity in a state $p_{+}^0$ as
$P_{+,+}^{\alpha,\beta}$, and the probability to  obtain $p_-^{\alpha,\beta}$
in a measurement $e_{\alpha,\beta}$ on an entity in a state $p_{+}^0$ as
$P_{+,-}^{\alpha,\beta}$. Analogously we define  $P_{-,+}^{\alpha,\beta}$ and
$P_{-,-}^{\alpha,\beta}$. We have: 
\beqn\label{eq:4a}
P_{+,+}^{\alpha,\beta}
&=&
|\langle\psi_+^0|\psi_+^{\alpha,\beta}\rangle|^2=cos^2{\beta\over 2}={1+cos\beta\over 2}
\\ 
\label{eq:4b}
P_{-,-}^{\alpha,\beta}
&=&
{1+cos\beta\over 2}
\\ \label{eq:4c} 
P_{+,-}^{\alpha,\beta}
&=&
P_{-,+}^{\alpha,\beta}=sin^2{\beta\over 2}={1-cos\beta\over 2}
\eeqn
Following eq.(\ref{eq:3}) we find that the set of states of a spin-${1\over 2}$
entity is given by: 
\beq\label{eq:5}
\Sigma_{1\over 2}
=\bigr\{p_+^{\alpha,\beta}|\alpha\in[0,2\pi],\beta\in
[0,\pi]\bigl\}
\eeq  

\subsubsection{Representation of the spin-1/2 states on ${\cal S}$.}   

Let ${\cal S}$ be a unit sphere in ${\Bbb R}^3$ with
its center in the origin. We represent every state
$p_+^{\alpha,\beta}\in\Sigma_{1\over 2}$ by the point in ${\cal S}$ with coordinates
$(cos\alpha sin\beta, sin\alpha sin\beta, cos\beta)$. It is clear (as a
consequence of the definition of the Euler angles), that the representation
of $\Sigma_{1\over 2}$ in ${\cal S}$ is one to one and onto.  The state $p_+^{\alpha,\beta}$
corresponds with the point on the sphere in the direction corresponding with the Euler angles   
$\alpha,\beta$, the state $p_-^{\alpha,\beta}$ corresponds with the antipodic point.

\subsubsection{The transition probability for coherent spin-S states.}   
 
We proceed along the same lines as in the previous section. We
denote a measurement characterized by the Euler angles $\alpha,\beta,\gamma$
as $e_{\alpha,\beta,\gamma}$. If the initial state of the entity corresponds
with a magnetic spin quantum number $M$ (which can be equal to $-S, -S+1, \dots, S-1, S$) we
denote it as
$p_M^0$. We represent $p_M^0$ by
the vector $\psi_M^0=(\delta_{M,i})_i\in{\Bbb C}^{2S+1}$,
where $i$ take values in $\{-S, -S+1, \dots, S-1, S\}$ (this indexation simplifies the
expression and will be used throughout the paper). The eigenvectors corresponding to a
measurement $e_{\alpha,\beta,\gamma}$ are again the same as the ones we obtain when
we rotate the initial states by an active rotation characterized by the Euler
angles $\alpha,\beta,\gamma$.  This active rotation is represented by a
unitary operator acting on ${\Bbb C}^{2S+1}$ that corresponds with the
following matrix (the explicit expression is again derived in \cite{wig59}):
\beq\label{wignerstate}
\hspace{-0.1cm}
M_{\alpha,\beta,\gamma}^S=\Biggl(
e^{-I j\alpha}e^{-I i\gamma}C_{i,j}^S\sum_k{(-1)^k
\bigl(cos{\beta\over 2}\bigr)^{2S+i-j-2k}\bigl(-sin{\beta\over 2}\bigr)^{j-i+2k}
\over(S-j-k)!(S+i-k)!(k+j-i)!k!}
\Biggr)_{i,j}
\eeq
(again $i$ and $j$ take values in $\{-S, -S+1, \dots, S-1, S\}$), where:
\beq
C_{i,j}^S=\sqrt{(S+i)!(S-i)!(S+j)!(S-j)!}
\eeq
and where the summation goes over all $k$ such that all exponents are non-negative.  This implies:
\beq
max\{0,M-M'\}\leq k\leq min\{S-M',S+M\}
\eeq
Thus, for the measurement $e_{\alpha,\beta,\gamma}$ we have the following collection of eigenvectors
that correspond with the possible outcome states $\{p_M^{\alpha,\beta}\}_M$,
labeled by the different possible values that can be taken by $M$ (again there is no
dependence of the states on $\gamma$):
\beq
\forall M: \psi_M^{\alpha,\beta,\gamma}=M_{\alpha,\beta,\gamma}^S\psi_M^0
\eeq
and explicitly we have:
\beq
\hspace{-0.1cm}
\psi_M^{\alpha,\beta,\gamma}=\Biggl(
e^{-I i\alpha}C_{M,i}^S\sum_k{(-1)^k
\bigl(cos{\beta\over 2}\bigr)^{2S+M-i-2k}\bigl(-sin{\beta\over 2}\bigr)^{i-M+2k}
\over(S-i-k)!(S+M-k)!(k+i-M)!k!}
\Biggr)_i e^{-I M\gamma}
\eeq
As a consequence, one easily verifies that we have the following relations between these states:
\beq \label{p-relation1}
p_{M}^{\alpha,\beta}=p_{-M}^{\alpha+\pi,\pi-\beta}
\eeq
and in particular:
\beq\label{p-relation2}
p_{0}^{\alpha,\beta}=p_{0}^{\alpha+\pi,\pi-\beta}
\eeq
Moreover, the above written equations are the only ones that relate identical states for different
values of $M$ and different directions characterized by the Euler angles.
We have the following transition probability for a
change of state of
$p_M^0$ into
$p_{M'}^{\alpha,\beta}$:
\beq\label{wigprob}
\hspace{-0.2cm}
P_{M,M'}^{\alpha,\beta}=|\langle \psi_M^0|\psi_{M'}^{\alpha,\beta}\rangle|^2=
\Biggl|
C_{M,M'}^S\sum_k{(-1)^k
\bigl(cos{\beta\over 2}\bigr)^{2S+M-M'-2k}\bigl(-sin{\beta\over 2}\bigr)^{M'-M+2k}
\over(S-M'-k)!(S+M-k)!(k+M'-M)!k!}
\Biggr|^2
\eeq
So again the transition probabilities depend only on the angle $\beta$.

\subsubsection{Representation of the spin-S states on ${\cal S}$.}\label{maj.rep.}   

As a first step in considering the spin-S entity as $2S$ individual spin-1/2 entities in a compound
system we introduce a representation for its states as proper states of the individual entities in
the compound system:
\par
\smallskip 
\noindent
{\it 
We will represent the state $p_M^{\alpha,\beta}$ of a spin-S entity 
by the following proper states of
the $2S$ individual spin-1/2 entities in the compound system:
}
\beq
\left\{ 
\begin{array}{l}
S+M\ {\it proper\ states}\ p_+^{\alpha,\beta}\\
S-M\ {\it proper\ states}\ p_-^{\alpha,\beta}
\end{array} 
\right\delimiter0
\eeq
\par
\smallskip 
\noindent
One easily verifies that, due to eq.(\ref{p-relation1}) and eq.(\ref{p-relation2}), this
representation is one to one.  
Since every spin-1/2 state corresponds with a point
on a sphere, we have represented the state $p_M^{\alpha,\beta}$ of a spin-S entity by $2S$
points on sphere:
$S+M$ points on the sphere in the direction corresponding with the Euler angles   
$\alpha,\beta$ and $S-M$ points in the opposite direction.  This is the Majorana representation for
the specific case of coherent spin-S states.  It is onto on the possible arrangements of $2S$ points
on a sphere which are such that all
points are in one of the two opposite locations on the sphere.  As a consequence, the representation
is completely determined by the angles $\alpha,\beta$ and the value of $M$.
 
\subsection{Relating the spin-S transition probabilities to the spin-1/2 transition
probabilities.}\label{sec:sym}

As already announced, in this section we show in which way the transition probability of spin-S states
can be related to the spin-1/2 transition probabilities.  Let us consider the
Majorana representation of a spin-S state $p_M^0$. The quantum description of a
compound system consisting of
$2S$ individual spin-1/2 entities, $S+M$ in a proper state $p_+^0$ and  $S-M$ in a proper state
$p_-^0$, can be written as the following product:
\beqa
&\underbrace{\psi_+^0\otimes\ldots\otimes \psi_+^0}\otimes\underbrace{\psi_-^0\otimes\ldots\otimes \psi_-^0}&\\
&\hspace{1.2cm} S+M \hspace{1.4cm} S-M \hspace{1.1cm}&
\eeqa
For this expression we introduce the following reduced notation:
\beq
(\otimes \psi_+^0)^{S+M}(\otimes \psi_-^0)^{S-M}
\eeq
A way to abstract from the ordering of the proper states of the individual spin-1/2
entities in the product is symmetrization (for the moment, we don't attach any physical significance to
this symmetrization). To do this, we have to sum over all possible permutations of the proper states
in the product. If we denote all distinguishable permutations\footnote{Permutations can be
indistinguishable if we permute identical elements.} of
$2S$ elements by
$\Pi$, we can write this as:
\beq\label{eq:sym}
\sum_{\pi\in\Pi}
\pi\bigl((\otimes \psi_+^0)^{S+M}(\otimes \psi_-^0)^{S-M}\bigr)
\eeq
Due to this symmetrization, we loose the normalization:
\beqa
\bigl|\sum_{\pi\in\Pi} 
\pi\bigl((\otimes \psi_+^0)^{S+M}(\otimes \psi_-^0)^{S-M}\bigr)\bigr|^2
&=&\bigl|\sum_{\pi,\pi'\in\Pi}
\pi\bigl((\otimes \psi_+^0)^{S+M}(\otimes \psi_-^0)^{S-M}\bigr)
\pi'\bigl(\overline{(\otimes \psi_+^0)^{S+M}(\otimes \psi_-^0)^{S-M}}\bigr)\bigr|\\
&=&\bigl|\sum_{\pi\in\Pi}1\bigr|={(2S)!\over(S+M)!(S-M)!}
\eeqa
since there are $(2S)!$ permutations of $2S$ elements of which $(S+M)!(S-M)!$ are
indistinguishable due to permutation of identical spin-1/2 states.  We introduce the following
abbreviation:
\beq
N=\sqrt{(2S)!\over(S+M)!(S-M)!}
\eeq
Let us calculate the the transition probability of the above constructed (normalized)
symmetrical superposition to a product state that corresponds with $S+M'$ individual spin-1/2 entities
in a proper state
$p_+^{\alpha,\beta}$ and 
$S-M'$ in a proper state $p_-^{\alpha,\beta}$:
\beqa
&&\hspace{-1.5cm}
\bigl|\bigl\langle
{1\over N}\sum_{\pi\in\Pi}
\pi\bigl((\otimes \psi_+^0)^{S+M}(\otimes \psi_-^0)^{S-M}\bigr)
\bigm|
(\otimes \psi_+^{\alpha,\beta})^{S+M'}(\otimes \psi_-^{\alpha,\beta})^{S-M'}
\bigr\rangle\bigr|^2
\\
&&\hspace{2.6cm}=
\bigl|{1\over N}\sum_{\pi\in\Pi}
\bigl\langle
\pi\bigl((\otimes \psi_+^0)^{S+M}(\otimes \psi_-^0)^{S-M}\bigr)
\bigm|
(\otimes \psi_+^{\alpha,\beta})^{S+M'}(\otimes \psi_-^{\alpha,\beta})^{S-M'}
\bigr\rangle\bigr|^2
\\
&&\hspace{2.6cm}=
\bigl|{1\over N}
\sum_k a_k
\langle\psi_+^0|\psi_+^{\alpha,\beta}\rangle^{b_{++}}
\langle\psi_-^0|\psi_+^{\alpha,\beta}\rangle^{b_{-+}}
\langle\psi_+^0|\psi_-^{\alpha,\beta}\rangle^{b_{+-}}
\langle\psi_-^0|\psi_-^{\alpha,\beta}\rangle^{b_{--}}
\bigr|^2
\eeqa
where the exponents have to fulfill the following equations:
\beqn
&&b_{++}+b_{+-}=S+M\\
&&b_{-+}+b_{--}=S-M\\
&&b_{++}+b_{-+}=S+M'\\
&&b_{+-}+b_{--}=S-M'
\eeqn
of which only three are independent (which justifies the summation over $k$).  We can
parameterize the solutions in the following way:
\beqn
&&b_{++}=S+M-k\\
&&b_{-+}=S+M'-b_{++}=M'-M+k\\
&&b_{+-}=S+M-b_{++}=k\\
&&b_{--}=S-M'-b+-=S-M'-k
\eeqn
Since all exponents should be non-negative, $k$ only takes integer values that fulfill:
\beq
max\{0,M-M'\}\leq k\leq min\{S-M',S+M\}
\eeq
The constants $a_k$ are equal to the number of permutations that lead to exponents that correspond
to this value of $k$, and thus they are the product of: 
\par
\smallskip 
\noindent
{\it 
{\bf 1.} the number of possibilities to select $b_{++}$ times the factor $\otimes\psi_+^{\alpha,\beta}$
out of the $S+M'$ ones
\par
\smallskip 
\noindent
{\bf 2.} the number of possibilities to select $b_{+-}$ times the factor $\otimes\psi_-^{\alpha,\beta}$
out of the $S-M'$ ones
}
\par
\smallskip 
\noindent
Thus we have:
\beq
a_k={(S+M')!\over (S+M-k)!(M'-M+k)!}{(S-M')!\over k!(S-M'-k)!}
\eeq
After substituting all this we obtain:
\beqa
&&\hspace{-1.0cm}
\bigl|\bigl\langle
{1\over N}\sum_{\pi\in\Pi}
\pi\bigl((\otimes \psi_+^0)^{S+M}(\otimes \psi_-^0)^{S-M}\bigr)
\bigm|
(\otimes \psi_+^{\alpha,\beta})^{S+M'}(\otimes \psi_-^{\alpha,\beta})^{S-M'}
\bigr\rangle\bigr|^2
\\
&&\hspace{-0cm}=
\Biggl|\sum_k
{\sqrt{(S+M)!}\sqrt{(S-M)!}(S+M')!(S-M')!
(-1)^k\bigl(cos{\beta\over 2}\bigr)^{2S+M-M'-2k}\bigl(-sin{\beta\over 2}\bigr)^{j-i+2k}
\over \sqrt{(2S)!}(S+M-k)!(M'-M+k)!k!(S-M'-k)!}
\Biggr|^2
\eeqa
This equation is equal to eq.(\ref{wigprob}) up to a constant:
\beq
\Biggl|\sqrt{(2S)!\over(S+M')!(S-M')!}\Biggr|^2={(2S)!\over(S+M')!(S-M')!}
\eeq
This are exactly the number of distinguishable orderings of the vectors in the product state:
\beqa
(\otimes \psi_+^{\alpha,\beta})^{S+M'}(\otimes \psi_-^{\alpha,\beta})^{S-M'}
\eeqa
i.e., all
possible representations of the Majorana representation for the outcome state
$p_{M'}^{\alpha,\beta}$ as a product state in $\bigl(\otimes{\Bbb C}^2\bigr)^{2S}$.  Thus
we are able to recover the spin-S transition probabilities in the tensor product space of the $2S$
Hilbert spaces related to the individual spin-1/2 entities which the compound system described in
$\bigl(\otimes{\Bbb C}^2\bigr)^{2S}$ consists of:
\beqa
\hspace{-0.8cm}\underbrace{|\langle\psi_M^0|
\psi_{M'}^{\alpha,\beta}\rangle|^2}
\hspace{-0.3cm}&=&\hspace{-0.3cm}
\underbrace{\sum_{\pi'\in\Pi}\bigl|\bigl\langle{1\over
N}\sum_{\pi\in\Pi}\pi\bigl((\otimes
\psi_+^0)^{S+M}(\otimes\psi_-^0)^{S-M}\bigr)\bigm|\pi'\bigl((\otimes \psi_+^{\alpha,\beta})^{S+M'}
(\otimes\psi_-^{\alpha,\beta})^{S-M'}\bigr)\bigr\rangle\bigr|^2}\\
\hspace{-0.3cm}
\in{\Bbb C}^{2S+1}\hspace{-0.2cm}&&\hspace{4.7cm}\in\bigl(\otimes{\Bbb
C}^2\bigr)^{2S}
\eeqa
\vspace{-3mm}
\par\noindent
or, when we denote by ${\cal H}_{M'}^{\alpha,\beta}$ the subspace of
$\bigl(\otimes{\Bbb C}^2\bigr)^{2S}$ spanned by the different vectors:
\beqa
\pi'\bigl((\otimes
\psi_+^{\alpha,\beta})^{S+M'}(\otimes\psi_-^{\alpha,\beta})^{S-M'}\bigr)
\eeqa
that correspond with the
different permutations $\pi'$:  
\beq
|\langle\psi_M^0| \psi_{M'}^{\alpha,\beta}\rangle|^2
=
\bigl|\bigl\langle{1\over N}\sum_{\pi\in\Pi}\pi\bigl((\otimes
\psi_+^0)^{S+M}(\otimes\psi_-^0)^{S-M}\bigr)\bigm|{\cal H}_{M'}^{\alpha,\beta}\bigr\rangle\bigr|^2
\eeq
where  $\langle\psi|{\cal H}_{M'}^{\alpha,\beta}\rangle$ stands for the projection of the
vector $\psi$ on the subspace ${\cal H}_{M'}^{\alpha,\beta}$. As a consequence we also have:
\beq
\sum_{M'}\bigl|\bigl\langle{1\over N}\sum_{\pi\in\Pi}\pi\bigl((\otimes
\psi_+^0)^{S+M}(\otimes\psi_-^0)^{S-M}\bigr)\bigm|{\cal H}_{M'}^{\alpha,\beta}\bigr\rangle\bigr|^2
=1
\eeq
We can summarize this as follows:
\par\medskip\noindent
{\it
Let: 
\beq
\left\delimiter0
\begin{array}{l}
\ \ (\otimes\psi_+^0)^{S+M}(\otimes\psi_-^0)^{S-M}\\
(\otimes \psi_+^{\alpha,\beta})^{S+M'}(\otimes\psi_-^{\alpha,\beta})^{S-M'}  
\end{array} 
\right\}
\in\bigl(\otimes{\Bbb C}^2\bigr)^{2S}
\eeq
respectively correspond with the Majorana representation of $\psi_M^0\in{\Bbb
C}^{2S+1}$ and of $\psi_{M'}^{\alpha,\beta}\in{\Bbb C}^{2S+1}$.  We realize the spin-S transition
probabilities in $\bigl(\otimes{\Bbb C}^2\bigr)^{2S}$ if we represent $\psi_M^0$ by:
\beq\label{sym.init.state}
{1\over N}\sum_{\pi\in\Pi}\pi\bigl((\otimes\psi_+^0)^{S+M}(\otimes\psi_-^0)^{S-M}\bigr)
\eeq
and $\psi_{M'}^{\alpha,\beta}$ by the subspace spanned by: 
\beq
\Bigl\{\pi\bigl((\otimes
\psi_+^{\alpha,\beta})^{S+M'}(\otimes\psi_-^{\alpha,\beta})^{S-M'}\bigr)
\bigm|\pi\in\Pi\Bigr\}
\eeq
}
\par\medskip\noindent
We also make the following important observation:
\par\smallskip\noindent
{\it
Due to the absence of an ordering for the proper states of the individual spin-1/2
entities that appear in the Majorana representation of a spin-S state, there is a one to one onto
correspondence between: 
\par
\smallskip 
\noindent
{\bf 1.}  the spin-S states $\psi_M^0\in{\Bbb C}^{2S+1}$ themselves
{\bf 2.} the symmetrized superpositions of $\bigl(\otimes{\Bbb C}^2\bigr)^{2S}$ given by
eq.(\ref{sym.init.state})
{\bf 3.} the subspaces ${\cal H}_{M}^0$ of $\bigl(\otimes{\Bbb
C}^2\bigr)^{2S}$ obtained through permutation of the order in the products
$(\otimes\psi_+^0)^{S+M}(\otimes\psi_-^0)^{S-M}$
}
\par
\smallskip 
\noindent
A straightforward consequence of this is that the direct sum $\oplus_M{\cal
H}_M^{\alpha,\beta}$ is exactly $\bigl(\otimes{\Bbb C}^2\bigr)^{2S}$
i.e., these collections of subspaces correspond in a one to one way with the orthonormal bases of
${\Bbb C}^{2S+1}$ that correspond with coherent spin-S states.  This
particular correspondence between the transition probability of a spin-S entity and
the transition probabilities of the $2S$ individual spin-1/2 entities in a compound quantum system
will enable us to represent measurements on a spin-S entity as
a measurement on each of the $2S$ individual entities in a compound system by applying the
construction of
\cite{HC:FPL}.

\subsection{Decomposition of a measurement on a spin-S entity in a measurement on each of the $2S$
individual spin-1/2 entities through the introduction of hidden correlations.}\label{rep.HC}

Also here we proceed in two steps, first we introduce hidden correlations on the spin-1/2 entities,
and then we try to get rid of the density matrices representing the proper states which do not
correspond with spin-1/2 states\footnote{By these spin-1/2 states we mean the so called 'pure
states'.}.

\subsubsection{Introduction of hidden correlations on the spin-1/2 entities.}

In section \ref{sec:sym} we have shown that we are able to obtain the transition probabilities
of a spin-S entity in the tensor product space $\bigl(\otimes{\Bbb
C}^2\bigr)^{2S}$, which is the quantum mechanical representation space for the description of
compound systems consisting of $2S$ individual spin-1/2 entities: we identify the initial
state of a measurement on the spin-S entity with a symmetrical superposition of the states of
the spin-1/2 entities in
its Majorana representation and the possible outcome states with the union of all possible product
states in
$\bigl(\otimes{\Bbb C}^2\bigr)^{2S}$ that correspond with different distinguishable orderings of
the spin-1/2 states in their respective Majorana representation.  This clearly allows us to apply
the representation for tensor product states by means of hidden correlations introduced in
\cite{HC:FPL}:
\par
\smallskip 
\noindent
{\it We say that there exist 'hidden correlations' between individual entities in a collection
$\{S_\nu\}_\nu$ if: 
{\bf 1.} a measurement on one individual entity $S_\alpha$ induces a change of the proper state of the
other individual entities
{\bf 2.} This change of proper state depends deterministically on the proper state transition of
$S_\alpha$ 
{\bf 3.} After this measurement, $S_\alpha$ cannot be influenced anymore by measurements on other
entities.}
\par
\smallskip 
\noindent
For compound quantum systems represented by a tensor product we obtained the following results:
\par
\smallskip 
\noindent
{\it Every compound quantum system $S$ consisting of a finite collection of
individual entities
$\{S_\nu\}_\nu$, and described by $\Psi_S\in\otimes_\nu{\cal H}_\nu$ has a hidden correlation
representation: {\bf 1.} Initially, every $S_\alpha$ is in a proper state $\omega_\alpha$. {\bf 2.} If
$S_\lambda$ takes the state
$\phi_\lambda$ due to a measurement on it after we have already performed
measurements on
$S_\alpha,\ldots,S_\kappa$, 
then the proper state of every not yet measured individual entity $S_\mu$ changes to 
$\omega_{\mu\circ\lambda\circ\kappa\circ\ldots\circ\alpha}$ which only depends on $\phi_\lambda$.  Both
$\omega_\alpha$ and
$\omega_{\mu\circ\lambda\circ\kappa\circ\ldots\circ\alpha}$, which are explicitly defined in
\cite{HC:FPL}, are represented by a density matrix which in general does not correspond
with a spin-1/2 state.}
\par
\smallskip 
\noindent
We have a compound system consisting of $2S$ individual spin-1/2 entities described
by eq.(\ref{sym.init.state})
and a measurement on this compound system of which an outcome is represented by:
\beq
\cup_{M'}\bigl\{\pi\bigl((\otimes
\psi_+^{\alpha,\beta})^{S+M'}(\otimes\psi_-^{\alpha,\beta})^{S-M'}\bigr)
\bigm|\pi\in\Pi\bigr\}
\eeq
where we identify all eigenvectors with the same value of $M'$ as the same outcome (since they
correspond with the same Majorana representation of a spin-S state).
Thus, we are able to represent every measurement on the spin-S entity as a
series of $2S$ consecutive measurements on the individual entities such that every
measurement on one of these entities induces a transition of the proper state of the not yet measured
entities.  There is one important feature of the representation introduced in
\cite{HC:FPL} which, at first sight poses some problems for the application within the context of
this paper.  Namely, the state transitions introduced by hidden correlation might change the initial
proper states which correspond in a one to one way with spin-1/2 states into proper states
represented by a density matrix which doesn't.  Nonetheless, as we show in the following section,
this situation can always be avoided.  In the case of maximal spin states, i.e.,
$|M|=S$, all proper states in the Majorana representation are the same and thus, the
representative symmetrical superposition is a product, which is its own biorthogonal decomposition:
\beq
(\otimes \psi_+^0)^{2S}=
\psi_+^0\otimes(\otimes \psi_+^0)^{2S-1}
\eeq 
According to the procedure outlined in \cite{HC:FPL}, we have to construct a map:
\beq
T:{\Bbb C}^2\rightarrow\bigl(\otimes{\Bbb C}^2\bigr)^{2S-1}
\eeq 
to obtain the transitions of the proper states of the individual entities when a measurement
on one of them gives an outcome state $p_+^{\alpha,\beta}$ or $p_-^{\alpha,\beta}$:
\beq
T(\psi_+^{\alpha,\beta})=T(\psi_-^{\alpha,\beta})=(\otimes \psi_+^0)^{2S-1}\\ 
\eeq
The proper state transition of one of the not yet measured individual entities can be found
through a map 
${\cal R}$ that relates these vectors $T(\psi_+^{\alpha,\beta})$ and $T(\psi_-^{\alpha,\beta})$ 
with a density matrix, by considering the coefficients in their biorthogonal decomposition (see
\cite{HC:FPL}).  We only find density matrices corresponding with $p_+^0$.  Since these were also the
initial proper states of these individual entities, they are not influenced by the
measurement.  Thus, the
$2S$ individual entities behave as a collection of indistinguishable but separated
spin-1/2 entities.

\subsubsection{Correspondence of the density matrices representing the proper states with spin-1/2
states.}

As mentioned above, due to the results of \cite{HC:FPL} one might think that one is forced to
consider density matrices for the proper states of the individual entities.  This is not true since we
only demand a global probabilistic correspondence with subsets of possible outcomes 
(namely the outcome
states corresponding with the vectors that span ${\cal H}_M^{\alpha,\beta}$ and not for every
individual entity in the compound systems).  In other words: {\it we are not able to
distinguish the different outcome states of the individual entities in the compound system, we only
can obtain some knowledge on how many are in a proper state represented by $\psi_+^{\alpha,\beta}$ and
how many are in a proper state represented by $\psi_-^{\alpha,\beta}$}.  As a consequence we have a
weaker constraint, which will allow us to decompose the density matrices into spin-1/2 states.  Due to
the results of \cite{hug93} and \cite{jay57} we know that every density matrix can easily be decomposed
in $2S$ 'pure' states ($S\geq 1$).  We illustrate this for the specific case of the initial
proper states.  If we have an initial state represented by eq.(\ref{sym.init.state}) we find the
proper states of the individual entities through a biorthogonal decomposition (see \cite{HC:FPL}):
\beq
\hspace{-0.8cm}
a_+\psi_+^0\otimes{1\over
N_+}\sum_{\pi\in\Pi}\pi\bigl((\otimes\psi_+^0)^{S+M-1}(\otimes\psi_-^0)^{S-M}\bigr) 
+
a_-\psi_-^0\otimes{1\over
N_-}\sum_{\pi\in\Pi}\pi\bigl((\otimes\psi_+^0)^{S+M}(\otimes\psi_-^0)^{S-M-1}\bigr)
\eeq
where:
\beqa
a_+&=&{N_+\over N}=\sqrt{(S+M)!(S-M)!\over(2S)!}\sqrt{(2S-1)!\over(S+M-1)!(S-M)!}
=\sqrt{S+M\over 2S}\\
a_-&=&{N_-\over N}=\sqrt{(S+M)!(S-M)!\over(2S)!}\sqrt{(2S-1)!\over(S+M)!(S-M-1)!}
=\sqrt{S-M\over 2S}
\eeqa
The corresponding density matrix $\omega_M$ is:
\beq
\pmatrix{
{S+M\over 2S} & 0\cr 
0  & {S-M\over 2S}\cr 
}
\eeq
in the base $\{\psi_+^0,\psi_-^0\}$.
Due to the symmetry of eq.(\ref{sym.init.state}), we obtain the same density matrix for every
individual entity.  Clearly, all these density matrices can be decomposed in $2S$ spin-1/2 states,
namely
$S+M$ times $\psi_+^0$ and $S-M$ times $\psi_-^0$.  Thus, as could be expected, the $2S$
identical density matrices $\omega_M$ are probabilistically equivalent with the spin-1/2 states in the
Majorana representation from which we started, since we cannot distinguish them.  In a similar way,
the other density matrices that appear in the procedure can be equivalently replaced by spin-1/2
states\footnote{Within the specific philosophical framework introduced in \cite{HC:FPL} we can
interpret this fact as an absence of hard acts of creation in the measurement on the spin-S
entity.  Actually, this is also in accordance with our main reason why we introduced hard acts of
creation, namely in order to characterize changes of the state space.}. An explicit realization of
this decomposition of density matrices for the specific case of coherent spin-1 states can be found in
\cite{coe95a}.

\section{Generalization to non-coherent states.}  

It is possible to extend the procedure of the previous sections to non-coherent spin states. 
Unfortunately in doing so, we loose the transparency of the procedure and also the explicit
reference to the spatial structure.
We proceed along the following steps: 
\par
\smallskip  
\noindent 
{\it 
{\bf 1.} For every spin-S state there exists a unique representation as $2S$ spin-1/2 states.  An
explicit procedure to do this is presented in the appendix at the end of this chapter.  Also the
outcome states of a measurement can be represented by their Majorana representation (as is done in
section \ref{coh.Maj.}).
{\bf 2.} We symmetrize the product state that corresponds to this $2S$ spin-1/2 states (that are
representative for the initial state) in the sense of section \ref{sec:sym}.  We obtain a
representation of the initial state in $\bigl(\otimes{\Bbb C}^2\bigr)^{2S}$. 
{\bf 3.} We apply the construction of \cite{HC:FPL} in order to introduce hidden correlations on the
$2S$ individual entities (in the sense of section \ref{rep.HC}).  In doing so, we obtain the
exact probabilities of a spin-S system when represented as a compound system consisting of
$2S$ individual spin-1/2 entities.}  
\par
\smallskip 
\noindent
Of course, this rather straightforward induction of the procedure presented in the previous
section is not yet an explicit proof. However, the explicit proof doesn't
contribute in any way to any deeper insights and is therefore omitted.
 
\section{A classical mechanistic representation in ${\Bbb R}^3$.}\label{sec:HC-HF}
 
In section 3 of \cite{HC:FPL} we have explained a procedure that enables the construction of
classical mechanistic representations for hidden correlation representations of general physical
entities.  For the spin-S entities introduced in this paper, we can realize such a
classical mechanistic representations easily:  
in \cite{aer86}, Aerts has been able to construct a very simple classical mechanistic model system 
(i.e., a model system in which appear only Kolmogorovian probability measures) in the three
dimensional space of reals for a measurement on a spin-1/2 entity with the states also represented on
the Poincar\'e sphere; thus, if we represent the measurements on the individual spin-1/2 entities that
appear in our representation by Aerts' model system, our representation gives rise to a classical
mechanistic model system in the three  dimensional space of reals for all spin-S entities.

\section{Conclusion.}

We are able to represent a measurement on a spin-S entity as a measurement on
each of $2S$ individual spin-1/2 entities in a compound system if we introduce hidden correlations,
i.e., a measurement on one of the individual entities induces a transition of the proper state
of the other ones.  If the spin-S entity is in a {\it maximal spin state} ($|M|=S$), the
$2S$ individual entities behave as a collection of indistinguishable but separated 
quantum entities.  If not so, the kind of correlations that we have to introduce are the same ones as
for compound quantum systems described in the tensor product by a symmetrical superposition.     
  
\section{Appendix: a Bacry-like procedure for the Majorana representation.}

Remarkably, although by leading authors in the field of angular momentum techniques it is claimed
that Bacry's proof is equivalent with Majorana's (see \cite{maj32}), it seems that this is not
completely true\footnote{We remark that this difference has lead the present author to the incorrect
remark that his representation of states of a spin-$1$ entity as two points on a sphere is differed
from the Majorana representation (see \cite{coe95a} footnote 2).}. Nonetheless, as we will show in
this appendix, Bacry's approach reveals a new way to 'generate' the Majorana representation when it is
adapted in a proper way to coincide with Majorana's representation.  

Bacry's procedure for representing spin-S states as $2S$ spin-1/2 states is the following.  Let
$\psi\in{\Bbb C}^{2S+1}$ be a vector representative for the spin-S state $p$, and let
$\{\psi_{-S},\ldots,\psi_S\}$ be the coordinates of $\psi$ in a given basis (we repeat that the
indices take values in $\{-S, -S+1, \dots, S-1, S\}$).  We can consider the following polynomial in
$x$:
\beq\label{poly} 
K_{p}^{Bacry}[x]
=\sum_i\psi_i x^{S-i}
\eeq
For every such polynomial with complex coefficients, there exists a factorization in factors of first
order in $x$, and these $2S$ factors are unique up to a multiplicative constant, i.e., they uniquely
determine a ray in ${\Bbb C}^2$.  Thus, every factor can be identified with one unique spin-1/2
state.  Since also the polynomial $K_{p}^{Bacry}[x]$ is determined up to a
multiplicative constant (because $\psi$ is) we can relate to the spin-S state $p$ $2S$ spin-1/2
states in a unique way.  Unfortunately, these spin-1/2 states are not the ones of the Majorana
representation which we discussed in section
\ref{maj.rep.}.  Consider the
spin-1/2 states in the majorana representation of a spin-S state $p_M^{\alpha,\beta}$. If we proceed
along the lines of the above mentioned procedure we find the following polynomial representative
for $p_+^{\alpha,\beta}$:
\beq
K_+^{\alpha,\beta}[x]=
e^{I{-\alpha-\gamma\over 2}}cos{\beta\over 2}x
+e^{I{\alpha-\gamma\over 2}}sin{\beta\over 2}
\eeq
and for $p_-^{\alpha,\beta}$:
\beq
K_-^{\alpha,\beta}[x]=
-e^{I{-\alpha+\gamma\over 2}}sin{\beta\over 2}x
+e^{I{\alpha+\gamma\over 2}}cos{\beta\over 2}
\eeq
Thus we find:
\beqa
K_{p_M^{\alpha,\beta}}^{Bacry}[x]
&=&
(K_+^{\alpha,\beta}[x])^{M+S}(K_-^{\alpha,\beta}[x])^{M-S}\\
&=& \sum_i\psi_i(p_M^{\alpha,\beta})x^{2S+1-i}
\eeqa
where we have (the computation proceeds along the same lines as in section \ref{sec:sym}):
\beq\label{bacryvect}
\hspace{-0.7cm}\psi_i(p_M^{\alpha,\beta})=
\sum_k{(-1)^k(S+M')!(S-M')!\bigl(cos{\beta\over 2}\bigr)^{2S+M-i-2k}\bigl(-sin{\beta\over
2}\bigr)^{i-M+2k}\over(S-i-k)!(S+M-k)!(k+i-M)!k!}
e^{-I (i\alpha+M\gamma)}
\eeq
and this leads to a vector $\bigl(\psi_i(p_M^{\alpha,\beta})\bigr)_i$ which defers definitely from
$\psi_M^{\alpha,\beta}$ given by eq.(\ref{wignerstate}).  Nonetheless, if in stead of considering the
polynomial $K_{p_M^{\alpha,\beta}}^{Bacry}[x]$ we consider a modified polynomial
$K_{p_M^{\alpha,\beta}}^{Majorana}[x]$, we do find the decomposition that corresponds with the
Majorana representation.  By comparing eq.(\ref{bacryvect}) and eq.(\ref{wignerstate}) one easily sees
that:
\beq
\psi_M^{\alpha,\beta}=\Biggl(\sqrt{(S+M)!(S-M)!\over
(S+i)!(S-i)!}\psi_i(p_M^{\alpha,\beta})\Biggr)_i
\eeq
Thus, for
$\psi\in{\Bbb C}^{2S+1}$, representative for the spin-S state $p$, we have to consider the
following polynomial in $x$:
\beq\label{poly}
K_{p}^{Majorana}[x]=
\sum_i\sqrt{(S+i)!(S-i)!}\,\psi_i\,x^{S-i}
\eeq
in stead of $K_{p}^{Bacry}[x]$ in order to obtain the Majorana representation.  Since this procedure
for finding a Majorana representation for spin-S states goes for any spin-S state (and not only
for coherent ones), we can use it to generalize our model introduced in section \ref{coh.Maj.} for
coherent spin-S states to non-coherent spin-S states.


\begin{thebibliography}{99}

\bibitem{aab94} 
\smallskip
\noindent {\bf T. Aaberge}, {\it Helv. Phys. Acta} {\bf 67}, 127 (1994).
\vspace{-0.2cm}

\bibitem{acc82}
\smallskip
{\bf L. Accardi}, {\it Nuovo Cimento} {\bf 34}, 161 (1982).  
\vspace{-0.2cm}

\bibitem{aer86}  
\smallskip
\noindent {\bf D. Aerts}, {\it J. Math. Phys.} {\bf 27}, 202 (1986). 
\vspace{-0.2cm}

\bibitem{bac74} 
\smallskip
\noindent {\bf H. Bacry}, {\it J. Math. Phys.} {\bf 15}, 1686 (1974).
\vspace{-0.2cm}

\bibitem{bie81}  
\smallskip
\noindent {\bf L.C. Biedenharn and J.D. Louck}, {\it Angular Momentum in Quantum Physics},
Addison-Wesley, Reading (1981).
\vspace{-0.2cm}

\bibitem{blo45}  
\smallskip
\noindent {\bf F. Bloch and I.I. Rabi}, {\it Rev. Mod. Phys.} {\bf 17}, 237 (1945).
\vspace{-0.2cm}

\bibitem{coe95a}  
\smallskip
\noindent {\bf B. Coecke}, {\it Helv. Phys. Acta} {\bf 68}, 396 (1995).
\vspace{-0.2cm}

\bibitem{HC:FPL}  
\smallskip
\noindent {\bf B. Coecke}, {\it Found. Phys.} {\bf 28}, 1109 (1998). 
\vspace{-0.2cm}

\bibitem{hug93} 
\smallskip
\noindent {\bf L.P. Hughston, R. Jozsa and W.K. Wooters}, {\it Phys. Lett.} {\bf 183A}, 14
(1993).
\vspace{-0.2cm}

\bibitem{jay57} 
\smallskip
\noindent {\bf E.T. Jaynes}, {\it Phys. Rev.} {\bf 108}, 171 (1957).
\vspace{-0.2cm}

\bibitem{maj32} 
\smallskip
\noindent {\bf E. Majorana}, {\it Nuovo Cimento} {\bf 9}, 43 (1932).
\vspace{-0.2cm}

\bibitem{mie68}
\smallskip
{\bf B. Mielnik}, {\it Comm. Math. Phys.} {\bf 9}, 55 (1968).
\vspace{-0.2cm}

\bibitem{pau}  
\smallskip
\noindent {\bf W. Pauli}, {\it Die Allgemeinen Prinzipien der Wellenmechanic}, Handbuch der Physik
Vol. V, Part I, Springer-Verlag, Berlin (1958).
\vspace{-0.2cm}

\bibitem{pir}  
\smallskip
\noindent {\bf C. Piron}, {\it Foundations of Quantum Physics}, W.A. Benjamin,
London (1976).
\vspace{-0.2cm}

\bibitem{sch77} 
\smallskip
\noindent {\bf J. Schwinger}, {\it Trans. N.Y. Acad. Sci.} [II] {\bf 38}, 170 (1977).
\vspace{-0.2cm}

\bibitem{wig59} 
\smallskip
\noindent {\bf E.P. Wigner}, {\it Group Theory and its Applications to the Quantum Mechanics of
Atomic Spectra}, Academic press, London (1959).

\end{thebibliography}
\end{document}